\documentclass[10pt,letterpaper,compsoc,conference]{iiswc23}

\usepackage{cite}
\usepackage{amsmath,amssymb,amsfonts}
\usepackage{algorithmic}
\usepackage{graphicx}
\usepackage[dvipsnames]{xcolor}
\usepackage[final]{microtype}
\usepackage[italic]{mathastext}
\usepackage{libertine}
\usepackage[T1]{fontenc}
\usepackage{textcomp}
\usepackage[varqu,varl]{zi4}
\usepackage[all]{nowidow}
\usepackage[auth-lg,affil-it]{authblk}
\usepackage[keeplastbox]{flushend}
\usepackage{hyperref}

\graphicspath{{figure/}}
\usepackage{url}

\usepackage{makecell}
\usepackage{colortbl}
\usepackage{tikz}
\usepackage{mathtools}
\usepackage{multirow}
\usepackage{fancyhdr}
\usepackage{soul}
\usepackage{subfig}
\usepackage{booktabs} % For formal tables
\usepackage{pifont}
\usepackage[T1]{fontenc}
\newcommand{\modelname}{MMBench }
\newcommand{\modelnamenospace}{MMBench}
\begin{document}

%% EDIT TITLE BELOW

% \title{Tools-MMBench: An End-to-End Tool for Analyzing the Hardware-Software Implications of Multi-modal DNN Workloads}

\title{MMBench: Benchmarking End-to-End Multi-modal DNNs and Understanding Their Hardware-Software Implications}

%% DO NOT EDIT THE FOLLOWING

\renewcommand\Authsep{\qquad}
\renewcommand\Authand{\qquad}
\renewcommand\Authands{\qquad}

%% EDIT AUTHOR LIST BELOW

% \author{Author1 Name}
% \author{Author2 Name}
% \author{Author3 Name}
% \affil{Full Name of Awesome School}

%% ALTERNATIVE FORMAT FOR MULTIPLE SCHOOLS:
%% 
\author[1]{Cheng Xu}
\author[1]{Xiaofeng Hou}
\author[1]{Jiacheng Liu}
\author[1]{Chao Li}
\author[1]{Tianhao Huang}
\author[1]{Xiaozhi Zhu}
\author[1]{\\Mo Niu}
\author[1]{Lingyu Sun}
\author[1]{Peng Tang}
\author[1]{Tongqiao Xu}
\author[2]{\\Kwang-Ting Cheng}
\author[1]{Minyi Guo}
\affil[1]{Shanghai Jiao Tong University}
\affil[2]{Hongkong University of Science and Technology}

\maketitle

%% EDIT YOUR PAPER'S CONTENTS BELOW

\begin{abstract}
The explosive growth of various types of big data and advances in AI technologies have catalyzed a new type of workloads called multi-modal DNNs. Multi-modal DNNs are capable of interpreting and reasoning about information from multiple modalities, making them more applicable to real-world AI scenarios. In recent research, multi-modal DNNs have outperformed the best uni-modal DNN in a wide range of distributed computing applications from traditional multimedia systems to emerging autonomous edge systems. However, despite their importance and superiority, very limited research attention has been devoted to understand the characteristics of multi-modal DNNs and their implications on current computing software/hardware platforms. 
Existing benchmarks either target uni-modal DNNs or only focus on the algorithm characteristics of multi-modal DNNs. 
There lacks representative benchmark suites that provide comprehensive system and architecture level analysis of multi-modal networks.

To advance the understanding of these multi-modal DNN workloads and facilitate related research, we present \modelnamenospace, an open-source, end-to-end benchmark suite consisting of a set of real-world multi-modal DNN workloads with relevant performance metrics for evaluation. We then use \modelname to conduct an in-depth analysis on the characteristics of multi-modal DNNs. 
We demonstrate their unique characteristics of clear multi-stage execution, frequent synchronization and high heterogeneity, which distinguish them from conventional uni-modal DNNs.
Finally, we conduct a case study and extend our benchmark to edge devices. We hope that our work can provide insights for future software/hardware design and optimization to underpin multi-modal DNNs on both cloud and edge computing platforms.
\end{abstract}

\section{Introduction}\label{sec:introduction}
Multi-modal deep neural networks (DNNs) have attracted significant attention~\cite{Zhang2020MultimodalIR, Wang2020WhatMT, baltruvsaitis2018multimodal} in recent years. By fusing information from a variety of modalities, they can provide higher prediction accuracy than the best traditional uni-modal DNNs~\cite{Zhang2020MultimodalIR, Liang2021MultiBenchMB,cal}. In fact, multi-modal DNNs have been shown to outperform the best uni-modal DNNs by 5\% - 30\% accuracy in many important application fields~\cite{Arevalo2017GatedMU}. Furthermore, the development of perception technology and AI accelerators has facilitated the deployment and development of multi-modal DNNs in a wide range of real-world applications from conventional multimedia to emerging autonomous systems. 

Despite their superiority in performance, multi-modal DNNs possess several unique characteristics, that have never been explored before and would pose new challenges to system and architecture designs previously applied to uni-modal DNNs. These characteristics include:
\begin{itemize}
    \item \textbf{Three-stage Execution Pattern:} Most multi-modal DNN applications follow a common three-stage execution pattern. In the first stage, known as \textit{encoder}, independent neural networks are utilized to translate input modalities to distinct representations that are suitable for machine learning. These representations are then fed to the second stage, known as \textit{fusion} where they are federated. Finally, the task-specific head network produces the final results in the third stage, known as \textit{head}. 
    The three stages are executed in serial, and each stage exhibits different execution and resource usage patterns.

    \item \textbf{Intra-network Heterogeneity:} A multi-modal DNN shows great intra-network heterogeneity due to the use of different encoder and fusion networks. The first stage inherently involves different networks and operators to process different modalities e.g. CNNs for image modality and RNN/Transformers for text modality. Additionally, in the fusion stage, different fusion methods can be applied to federate the features of different modalities for different accuracy targets. As a result, there are no universal architectural solutions to optimize all modalities and stages, which are often dominated by heterogeneous operations. 

    \item \textbf{Frequent Synchronization:} The fusion stage in multi-modal DNNs incurs substantial synchronization operations compared to traditional uni-modal networks. In this stage, the fusion network waits for the completion of all modalities, and additional CPU-GPU synchronization is needed to process intermediate data, such as the feature maps generated from various modalities. These frequent synchronization operations can become a key performance bottleneck for multi-modal DNN computation, as they add extra latency and overhead.

\end{itemize}

As multi-modal DNNs become increasingly popular and differ significantly from conventional uni-modal DNNs, it is crucial to understand these unique characteristics and their implications for system and architecture designs. 
It is preferable to analyze their features and assign agile management strategies to maximize overall efficiency~\cite{sc}.
However, there is currently a lack of a well-designed benchmark suite that provides system- and architecture-level characterization of multi-modal DNNs. On one hand, uni-modal DNN benchmarks in previous studies~\cite{Re2020MLPerfIB, Hadidi2019CharacterizingTD, Gupta2020TheAI} cannot be directly applied to multi-modal DNNs due to the differences in their characteristics. On the other hand, existing multi-modal DNN benchmarks~\cite{Liang2021MultiBenchMB} only focus on algorithm-level features such as accuracy, model complexity and robustness without providing any analysis of system and architecture. Therefore, there is a strong motivation to develop benchmarks/tools specialized for multi-modal DNN applications and to explore their implications on today's computing architecture and systems.

In this paper, we propose \modelnamenospace, an end-to-end benchmark suite for multi-modal DNN applications. \modelname covers a wide spectrum of representative multi-modal applications across multiple major research areas. We also leverage \modelname to study the characteristics of multi-modal DNNs and their implications across the execution stacks. To the best of our knowledge, \modelname is the first benchmark suite specialized in architecture and system research in multi-modal computing. We design \modelname with the following principles:

\begin{itemize}
    % 涵盖应用范围广
    \item \textbf{Representativeness.} We construct \modelname using 9 end-to-end multi-modal DNN workloads from five of the most representative application domains, which cover traditional applications like multimedia and emerging domains such as autonomous driving. This approach ensures that \modelname is representative of multi-modal applications in use today.

    % 包含的网络特征多样性
    \item \textbf{Thoroughness.} We ensure that properties such as modality types, fusion methods, network structures in \modelname are diverse and cover a wide range of multi-modal DNNs in different domains. This level of thoroughness provides researchers with a detailed understanding of performance and potential areas for improvement for multi-modal DNNs.

    \item \textbf{Comprehensiveness.} 
    At \modelname, we have gone beyond offering just operational workloads and result scoreboards. We also provide comprehensive profiling tools and insights at the architecture and system levels. This level of support enables researchers to build on our work and advance the state of the art in multi-modal computing.
    
\end{itemize}

The rest of the paper is organized as follows. First, Section \ref{sec:background} provides the background and related work. Section \ref{sec:mmbench_design} details the designs of \modelnamenospace. Section \ref{sec:evaluation} shows the experimental methodologies and highlights the key features of multi-modal DNNs and their hardware-software implications. Section \ref{sec:case_study} gives two case studies that demonstrate how \modelname guides the system and architecture designs. Finally, Section \ref{sec:conclusion} concludes this paper.

\section{Background and Related Work}\label{sec:background}

% --------------------------------
\begin{figure}[t]
	\centering
	\includegraphics[width=0.98\linewidth]{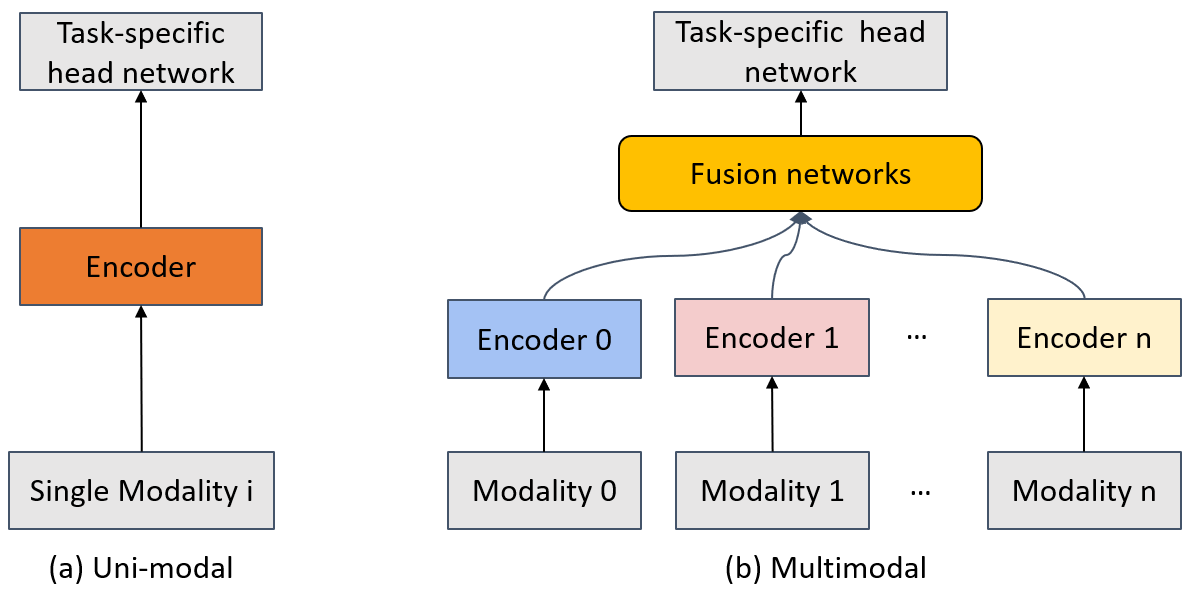}
	\caption{Schematic diagram of multi-modal and uni-modal network structures.} %----------------
	\label{fig:overview}
\end{figure}

\subsection{Basics of Multi-modal DNNs}
Multi-modal DNN is a kind of neural network that learns and improves through the experience of data from multiple modalities.
Figure \ref{fig:overview} shows the common structure of a multi-modal DNN compared to a uni-modal DNN. 
At a higher level, the multi-modal DNNs fuse the features from multiple modalities to produce more accurate predictions. 
Specifically, it consists of three main stages. In the first stage, input modalities are transferred to distinct representations suitable for machine learning by various representation learning methods such as CNNs. 
In the second stage, it leverages a fusion model to generate the multi-modal representation by federating these processed uni-modal representations. 
Finally, the multi-modal representation is fed into the task-specific network to produce the final prediction.

\begin{table}[t]
\caption{Commonly used fusion operators~\cite{Yin2022BMNASBM,Xu2021MUFASAMF}.}\label{tab:fusion}
\resizebox{\linewidth}{!}{
\begin{tabular}{|l|l|l|}
\hline
\textbf{Fusion type} & \textbf{Formulation of $F(x,y)$} & \textbf{Meaning}  \\ \hline
Zero  &     $0$    &    Discards these features         \\ \hline
Sum  &   $x + y$      &  Sum features   \\ \hline
Concat &  $\mathrm{ReLU} (\mathrm{Concat}(x, y) W + b)$& Concat features\\ \hline
Tensor  & $x\otimes y$ & Outer product-based attention\\ \hline
Attention   & $\mathrm{Softmax}(\frac{x y^T}{\sqrt{C}}y)$  & Use attention mechanism  \\ \hline
LinearGLU  & \begin{tabular}[c]{@{}l@{}}$\mathrm{GLU}(x W_1, y W_2)$\\ $ = x W_1 \odot \mathrm{Sigmoid}(y W_2)$\end{tabular} & linear layer with the GLU \\ \hline
\end{tabular}
}
\end{table}

Multi-modal DNNs have been demonstrated to outperform the uni-modal ones in various application fields~\cite{isca,Liang2021MultiBenchMB, Sun2020LearningRB}. 
Most of the current studies employ pretrained DNN
backbone models as modality encoders and mainly focus on finding more effective fusion or representation methods of different modalities~\cite{Zadeh2017TensorFN, Wang2020WhatMT, Zhang2020MultimodalIR}.
Commonly adopted fusion operators are presented in Table~\ref{tab:fusion}. 
Besides, the fusion technique can further be categorized into two main classes, namely early fusion methods~\cite{Zhou2008FeatureFO} and late fusion methods~\cite{Wu2004OptimalMF,bach2004multiple} depending on the depth of encoders to execute before fusion. 
Among these methods, Zhou et al ~\cite{Zhou2008FeatureFO} used a multiple discriminant analysis scheme to implement an early fusion approach that concatenates different modality features. 
Uperkernel learning~\cite{Wu2004OptimalMF} is the representative method of late fusion. 
Recently,  motivated by the ability of transformers~\cite{vaswani2017attention}, 
a branch of works use multi-modal transformers to model different modalities~\cite{Xu2022MultimodalLW}.

With the significant performance advance, mutli-modal DNNs have been widely adopted in various scenarios. Thus it is in urgent need of benchmarking multi-modal DNNs from system and architectural level to facilitate the optimization in their deployment. 

\subsection{Benchmarking Conventional DNNs}
Previous researches have paid extensive attention to characterizing features of uni-modal DNN applications~\cite{Xi2020SMAUG, Wu2019AccelergyAA, Kandiah2021AccelWattchAP, Wudenhe2021TPUPointAC}. 
We can broadly classify these works into three types according to their associated evaluation metrics: algorithm-oriented DNN benchmarks, architecture-oriented DNN benchmarks and DNN simulation frameworks.
Among them, algorithm-oriented DNN benchmarks~\cite{Re2020MLPerfIB, Coleman2017DAWNBenchA, Schefke2020DeepBenchOT, Adolf2016FathomRW} strive to incorporate and build a collection of representative DNN models to empower the performance and accuracy comparison of different DNN training and inference. 
Architecture-oriented DNN benchmarks~\cite{Zhang2022ACB, Almeida2019EmBenchQP, Zhang2018ASO} target on analyzing the architectural features of DNNs on computing systems of different sizes.
MLPerf~\cite{Re2020MLPerfIB} is a comprehensive benchmark for measuring ML inference performance across a spectrum of use cases. 
MDLBench~\cite{Zhang2022ACB}, Embench~\cite{Almeida2019EmBenchQP} and AIoTBench~\cite{Luo2020ComparisonAB} are representative benchmarks that characterize the features of different AI models on edge or mobile devices while NNBench-X~\cite{Xie2019NNBenchXBA}, GNNMark~\cite{Baruah2021GNNMarkAB} target on acceleration hardware design for different DNNs. 

However, the system and architecture level implications drawn by uni-modal DNN benchmarks can not be directly applied to multi-modal DNNs. Compared with uni-modal DNNs, multi-modal DNNs possess several unique characteristics such as clear stage divisions, frequent synchronization and high workload heterogeneity~\cite{cal}. There lack of specialized architecture-oriented multi-modal benchmarks.

\subsection{Benchmarking Multi-modal DNNs}
Some efforts have also been made in benchmarking the emerging workload of multi-modal DNNs. MultiBench~\cite{Liang2021MultiBenchMB} is a well-known benchmark suite in multi-modal algorithm research. MultiBench implements a wide spectrum of multi-modal applications and provides a reliable way to evaluate the performance across domains and modalities, the complexity during training/inference and the robustness to noisy and missing modalities. 
However, it only evaluates multi-modal DNNs from the algorithm aspect and does not provide system and architecture-level insights. Besides, MultiBench does not provide end-to-end implementation, which makes it insufficient to support architecture research especially for edge devices where raw data are collected from sensors and processed locally. 

Architecture benchmarks such as MLPerf~\cite{Re2020MLPerfIB}, DAWNBench~\cite{Coleman2017DAWNBenchA}, and AI-Benchmark~\cite{Gao2019AIBenchAI} can be applied to benchmarking multi-modal DNNs but requires significant modifications. MLPerf provides a comprehensive analysis of how fast systems can train and inference to a target quality metric while covering a wide range of models and areas. It measures a wide range of metrics such as training time, training cost, inference latency, and inference cost. However, as general-purpose full-system benchmarks, they lack specialized algorithm awareness and corresponding analysis in this emerging area. 
Specialized multi-modal benchmarks can better help design and deploy efficient multi-modal DNN systems.

\subsection{The Missing Piece of Multi-modal Research}
Multi-modal DNNs are applied in a wide range of applications in different fields.
Generally, multi-modal DNNs are more computing-intensive compared with traditional uni-modal DNNs, which may possibly lead to the problem of QoS and power budget violation~\cite{icpp1,icpp2}.
In data centers, we need to analyze complex multi-modal data for the highest algorithm performance (e.g., image- and text-based intelligence applications); in edge devices, we need to process raw data collected by multiple sensors locally with limited computational resources within QoS (e.g., autonomous driving). 
Supporting the inference of such diverse and heterogeneous workloads with high energy efficiency and low latency is becoming a great challenge.

\begin{table}[]
\scriptsize
\centering
\renewcommand\arraystretch{1.0}
\caption{Comparison of \modelname and other benchmarks~\cite{Liang2021MultiBenchMB,Re2020MLPerfIB,Coleman2017DAWNBenchA,Gao2019AIBenchAI}. \textit{H} refers to hardware, \textit{Ar} refers to architecture, \textit{S} refers to system, \textit{Al} refers to algorithm.}
\label{tab:comparison}
 \resizebox{\columnwidth}{!}{%
\begin{tabular}{|c|ccc|cc|}
\hline
\multicolumn{1}{|c|}{{\color[HTML]{000000} }}                                                                                    & \multicolumn{3}{c|}{\textbf{Uni-modal DNN}}                                                                                            & \multicolumn{2}{c|}{\textbf{Multi-modal DNN}}                                                                  \\ \cline{2-6} 
\multicolumn{1}{|c|}{\multirow{-2}{*}{{\color[HTML]{000000} \textbf{\begin{tabular}[c]{@{}c@{}}Benchmarks\end{tabular}}}}} & \multicolumn{1}{c|}{\textbf{MLPerf}} & \multicolumn{1}{c|}{\textbf{DAWNBench}}                                  & \textbf{AIBench} & \multicolumn{1}{c|}{\textbf{MultiBench}} & \textbf{Ours}                                                   \\ \hline
\textbf{Applications}                                                                                                                  & \multicolumn{1}{c|}{5}               & \multicolumn{1}{c|}{3}                                                   & 10               & \multicolumn{1}{c|}{15}                  & 9                                                               \\ \hline
\textbf{Objectives}                                                                                                              & \multicolumn{1}{c|}{H}              & \multicolumn{1}{c|}{\begin{tabular}[c]{@{}l@{}}H/Ar\end{tabular}} & H               & \multicolumn{1}{c|}{Al}               & \begin{tabular}[c]{@{}l@{}}H/Ar,\\S/Al\end{tabular} \\ \hline
\textbf{Cloud}                                                                                                                   & \multicolumn{1}{c|}{\checkmark}      & \multicolumn{1}{c|}{\checkmark}                                          & \checkmark       & \multicolumn{1}{c|}{\checkmark}          & \checkmark                                                      \\ \hline
\textbf{Edge}                                                                                                                    & \multicolumn{1}{c|}{\checkmark}      & \multicolumn{1}{c|}{x}                                                   & x                & \multicolumn{1}{c|}{x}                   & \checkmark                                                      \\ \hline
\textbf{End-to-End}                                                                                                                     & \multicolumn{1}{c|}{x}               & \multicolumn{1}{c|}{\checkmark}                                          & \checkmark       & \multicolumn{1}{c|}{x}                   & \checkmark                                                      \\ \hline
\textbf{Easy-to-Use}                                                                                                                & \multicolumn{1}{c|}{x}               & \multicolumn{1}{c|}{x}                                                   & x                & \multicolumn{1}{c|}{x}                   & \checkmark                                                      \\ \hline
\end{tabular}
}
\end{table}

In order to  support efficient reasoning on multi-modal networks in data centers and edge devices, there is an urgent need for  benchmarks that can accurately model the system and architecture level characteristics of multi-modal networks. However, there have been no such well-designed benchmark suites as presented in table~\ref{tab:comparison}. On one hand, the architecture-oriented DNN benchmarks have not covered this emerging research area yet. The implications of uni-modal DNNs can not be directly applied to multi-modal DNNs. On the other hand, existing multi-modal benchmarks all focus on analyzing algorithm-level characteristics. Thus, we present \modelname in this paper to bridge this gap and benefit further research in this area.

\begin{table*}[t]
\scriptsize
\centering
\renewcommand\arraystretch{1.0}
\caption{Characteristics of each applications in \modelnamenospace}
\label{tab:app}
 \resizebox{\linewidth}{!}{%
\begin{tabular}{|c|cc|cc|cc|cc|c|}
\hline
\textbf{\begin{tabular}[c]{@{}c@{}}Application\\ domain\end{tabular}} & \multicolumn{2}{c|}{\textbf{\begin{tabular}[c]{@{}c@{}}Multimedia\\ Application\end{tabular}}} & \multicolumn{2}{c|}{\textbf{\begin{tabular}[c]{@{}c@{}}Affective\\ Computing\end{tabular}}} & \multicolumn{2}{c|}{\textbf{\begin{tabular}[c]{@{}c@{}}Intelligent\\ Medicine\end{tabular}}} & \multicolumn{2}{c|}{\textbf{\begin{tabular}[c]{@{}c@{}}Smart\\ Robotics\end{tabular}}} & \textbf{\begin{tabular}[c]{@{}c@{}}Automatic\\ Driving\end{tabular}} \\ \hline
\textbf{Workload} & \multicolumn{1}{c|}{\textit{AV-mnist}} & \textit{MM-imdb} & \multicolumn{1}{c|}{\textit{CMU-mosei}} & \textit{MUStARD} & \multicolumn{1}{c|}{\textit{\textit{Medical VQA}}} & \textit{Medical Seg.} & \multicolumn{1}{c|}{\textit{Mujoco Push}} & \textit{Vision \& Touch} & \textit{TransFuser} \\ \hline
\textbf{Model size} & \multicolumn{1}{c|}{Small} & Large & \multicolumn{1}{c|}{Large} & Large & \multicolumn{1}{c|}{Large} & Medium & \multicolumn{1}{c|}{Medium} & Medium & Medium \\ \hline
\textbf{Modalties} & \multicolumn{1}{c|}{\textit{\begin{tabular}[c]{@{}c@{}}1.image,\\ 2.audio\end{tabular}}} & \textit{\begin{tabular}[c]{@{}c@{}}1.image,\\ 2.text\end{tabular}} & \multicolumn{1}{c|}{\textit{\begin{tabular}[c]{@{}c@{}}1.language,\\ 2.vision,\\ 3.audio\end{tabular}}} & \textit{\begin{tabular}[c]{@{}c@{}}1.language,\\ 2.vision,\\ 3.audio\end{tabular}} & \multicolumn{1}{c|}{\textit{\begin{tabular}[c]{@{}c@{}}1.image,\\ 2.text\end{tabular}}} & \textit{\begin{tabular}[c]{@{}c@{}}MRI scans\\ (T1, T1c, \\ T2, Flair)\end{tabular}} & \multicolumn{1}{c|}{\textit{\begin{tabular}[c]{@{}c@{}}1.position, \\ 2.sensor,\\ 3.image, 4.control\end{tabular}}} & \textit{\begin{tabular}[c]{@{}c@{}}1.image, 2.force,\\ 3.proprioception,\\ 4.depth\end{tabular}} & \textit{\begin{tabular}[c]{@{}c@{}}1.image\\ 2.LiDAR\end{tabular}} \\ \hline
\textbf{Encoders} & \multicolumn{1}{c|}{\textit{1,2: LeNet}} & \textit{\begin{tabular}[c]{@{}c@{}}1: VGG\\ 2: Albert\end{tabular}} & \multicolumn{1}{c|}{\textit{\begin{tabular}[c]{@{}c@{}}1: BERT\\ 2: OpenFace\\ 3: Librosa\end{tabular}}} & \textit{\begin{tabular}[c]{@{}c@{}}1: BERT\\ 2: OpenFace\\ 3: Librosa\end{tabular}} & \multicolumn{1}{c|}{\textit{\begin{tabular}[c]{@{}c@{}}1: DenseNet\\ 2: Roberta\end{tabular}}} & \textit{All: U-Net} & \multicolumn{1}{c|}{\textit{\begin{tabular}[c]{@{}c@{}}1,2,4: MLP\\ 3: CNN\end{tabular}}} & \textit{\begin{tabular}[c]{@{}c@{}}1,2,4: CNN\\ 3: MLP\end{tabular}} & \textit{1,2: ResNet} \\ \hline
\textbf{\begin{tabular}[c]{@{}c@{}}Fusion\\ methods\end{tabular}} & \multicolumn{1}{c|}{\textit{\begin{tabular}[c]{@{}c@{}}Concate,\\ tensor\end{tabular}}} & \textit{\begin{tabular}[c]{@{}c@{}}Concate,\\ tensor\end{tabular}} & \multicolumn{1}{c|}{\textit{\begin{tabular}[c]{@{}c@{}}Concate, \\ tensor, \\ transformer\end{tabular}}} & \textit{\begin{tabular}[c]{@{}c@{}}Concate, \\ tensor, \\ transformer\end{tabular}} & \multicolumn{1}{c|}{\textit{Transformer}} & \textit{Transformer} & \multicolumn{1}{c|}{\textit{\begin{tabular}[c]{@{}c@{}}Concate, \\ tensor, \\ transformer\end{tabular}}} & \textit{\begin{tabular}[c]{@{}c@{}}Concate,\\ tensor\end{tabular}} & \textit{transformer} \\ \hline
\textbf{Task} & \multicolumn{1}{c|}{Class.} & Class. & \multicolumn{1}{c|}{Reg.} & Class. & \multicolumn{1}{c|}{Gen.} & Seg. & \multicolumn{1}{c|}{Class.} & Class. & Class. \\ \hline
\end{tabular}
}
\end{table*}
% --------------

\section{The \modelname Suite}
\label{sec:mmbench_design}
In this section, we introduce the unique features of \modelname and the specific benchmark setup, i.e. the workloads and the profiling pipeline.

\subsection{Key Features of \modelnamenospace}
Besides the general design principles, \modelname possess the following unique features closely related with the characteristics of multi-modal DNNs, which distinguishes it from general-purpose benchmarks in this specific area: 
\begin{itemize}

    \item \textbf{Fine-grained Network Characterization.}
    From network structure level, multi-modal DNNs can be viewed as the assembly of multiple encoder networks, fusion network and head network, which require more fine-grained workload characterization~\cite{Yin2022BMNASBM,Liang2021MultiBenchMB}.
    It is inaccurate to use the average or max value of the entire application to characterize multi-modal DNNs since these sub-nets may greatly differ in execution pattern and resource usage. 
    \modelname provides options to split the multi-modal DNN into different stages and characterize the sub-nets respectively.
    
    \item \textbf{End-to-End Application Execution.} 
    From application level, the processing of raw data for multi-modal DNNs is time-consuming and often require end-to-end execution in real-world scenarios~\cite{ete1,Prakash2021MultiModalFT}.
    Many networks take processed data as input~\cite{Vinyals2017ShowAT,Castro2019TowardsMS}. Existing algorithm-oriented benchmarks tend to ignore these preprocessing  and provide links to the processed data~\cite{Liang2021MultiBenchMB}. 
    Ignoring the preprocessing part results in bias on the computing process.
    \modelname provides an end-to-end multi-modal processing benchmark that can help us understand the full computational process of multi-modal networks. 
    
    \item \textbf{User-friendly Profiler Integration.} 
    From the architecture profiler level, it is often unnecessary and time-consuming to utilize the entire dataset. A dataset-free computation abstraction can significantly ease the profiler usage.
    Many datasets in multi-modal neural network research are not open-sourced and require a lengthy application process. Besides, some datasets can take up to hundreds of Gigabyte~\cite{Prakash2021MultiModalFT}.
    \modelname still provides models and links to the datasets to prove that all applications are of high performance. However, \modelname also provides the option to abstract the computation when network accuracy is not needed.
    It can randomly generate the input with the same shape as the datasets, which allows computer architecture researchers to skip the tedious work of downloading and storing data and more easily analyze the system and architecture characteristics of multi-modal applications.
    Besides, popular accelerator simulation frameworks such as timeloop~\cite{Parashar2019TimeloopAS} simply take the data shape and network shape as input and outputs the latency and energy consumption. 
    \modelname is able to directly provide this abstraction and free users of manual conversion in the simulation.
\end{itemize}

\subsection{Applications in \modelnamenospace}
\modelname includes nine applications from the five most important multi-modal research domains ~\cite{Liang2021MultiBenchMB}. 
For the majority the applications, \modelname provides multiple fusion options covering popular fusion operators~\cite{Xu2021MUFASAMF,Yin2022BMNASBM}. 
\modelname implements all the applications in SOTA methods to ensure the practical value. 
The detail of these applications are presented in Table \ref{tab:app}. 

\textbf{Multimedia Application:} 
With the development of the internet, multimedia data (language, image, video, and audio) is becoming the largest source of the big data. 
\modelname rebuilds two of the most representative multimedia applications: (1) \textit{AV-mnist}~\cite{pham2019found} is assembled from images of handwritten digits and audio samples of spoken digits.   (2) \textit{MM-imdb}~\cite{Arevalo2017GatedMU} uses movie titles, metadata, and movie posters to perform multi-label classification of movie genres.
We rebuild these applications based on the implementation of \textit{MultiBench}. 
To make these applications end-to-end and represent the state-of-the-art performance, we replace the fragmented image processing pipeline with an end-to-end \textit{VGG} network~\cite{Simonyan2015VeryDC}, and use the pre-trained \textit{ALBERT} model~\cite{Lan2020ALBERTAL} to extract text features.

\textbf{Affective Computing:} 
Affect computing is the field that studies the perception of human affective states (emotions, sentiment, and personalities) from our natural display of multi-modal signals~\cite{Liang2021MultiBenchMB} 
including spanning language (spoken words), visual (facial expressions, gestures), and acoustic (prosody, speech tone)~\cite{Liang2021MultiBenchMB}. 
\modelname selects two of the most representative affective computing dataset that involving language, video and audio. (1) \textit{MUStARD} is a video corpus used in sarcasm discovery~\cite{Castro2019TowardsMS}. (2) \textit{CMU-mosei} is the largest dataset of  sentence-level sentiment analysis and emotion recognition in real-world online videos~\cite{Vinyals2017ShowAT}. We rebuild these applications from the original data to make it an end-to-end application. We  use \textit{MMSA-FET}~\cite{mao2022m} to extract features, and including all the modules in the forward pass of the data.

\textbf{Intelligent Medicine:} Modern medical decision-making often involves integrating complementary information from several sources such as lab tests, imaging reports, and patient-doctor conversations. 
Multi-modal DNNs can help doctors make sense of high-dimensional data and assist them in the diagnosis process. 
We build this workload based on \textit{ViLMedic}~\cite{delbrouck-etal-2022-vilmedic}, 
a vision-and-language medical library.
We also consider a multi-modal segmentation task that can accurately segment brain tumor from Magnetic Resonance Imaging (MRI) ~\cite{Zhang2022mmFormerMM}. We rebuild these applications to make them easy to profile 
with the standard profiling tools.

% TODO 这个要重写
\textbf{Smart Robotics:} 
Robotics is also a very important example of multi-modal computing. 
In order to achieve accurate control of robots, we add many sensors to them and collect multi-modal information (e.g., visual, force etc.). 
The decision making based on this multi-modal information requires the use of multi-modal networks. 
\modelname includes two representative robotic tasks: (1) \textit{Mujoco Push}~\cite{Lee2020MultimodalSF}, the goal of which is to predict the pose of the object being pushed by the robot end-effector using the collected multi-modal information, i.e., visual (RGB and depth), force, and proprioception sensors. (2) \textit{Vision \& Touch}~\cite{Lee2020MakingSO}, which aims to predict action-conditional learning objectives that capture forward dynamics of the different modalities.
We rebuild these workloads to add hooks to profile the different part of the neural network.

\textbf{Automatic Driving:} 
Automatic Driving normally refers to self-driving vehicles that move without the intervention of a human driver. 
An autonomous driving systems typically come equipped with both cameras and LiDAR sensors. In \modelname, we modify \textit{TransFuser}~\cite{Prakash2021MultiModalFT} which is an architecture for end-to-end driving with two main components: (1) a Multi-Modal Fusion Transformer for integrating information from multiple modalities (single-view image and LiDAR), and (2) an auto-regressive waypoint prediction network. 
To ease the usage, we extract the TransFuser network and free its dependency on the CARLA simulator~\cite{Dosovitskiy17}.

\begin{figure}[t]
	\centering
	\includegraphics[width=\linewidth]{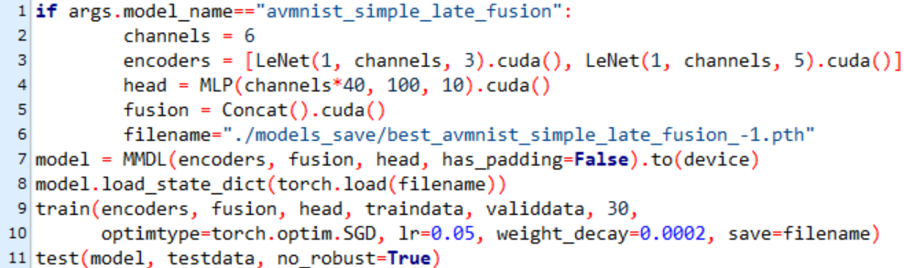}
	\caption{The code snippet of a standard multi-modal application implementation in \modelnamenospace.} %----------------
	\label{fig:code}
\end{figure}

\subsection{Implementation Details}
\begin{figure}[t]
	\centering
	\includegraphics[width=\linewidth]{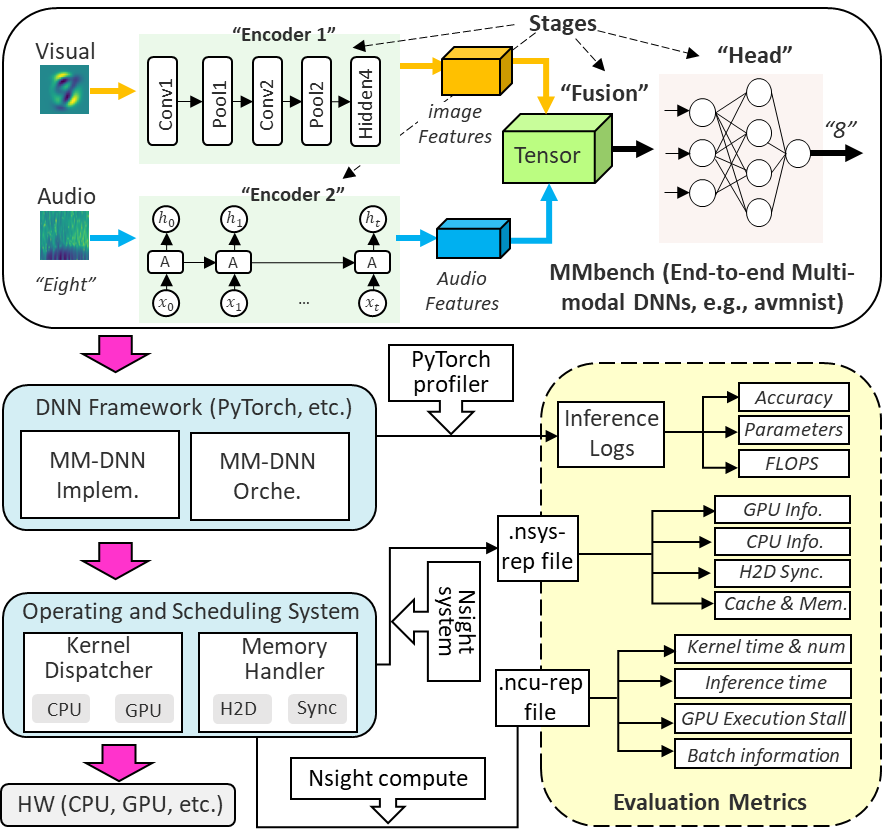}
	\caption{Profiling pipeline in \modelnamenospace.} %----------------
	\label{fig:profile}
\end{figure}

The entire \modelname is implemented in PyTorch, while different applications may possess their own dependencies. \modelname provide rich interfaces to enable users to control the workloads. Figure~\ref{fig:profile} presents a standard \modelname implementation. For a multi-modal application in \modelnamenospace, it applies specific encoder networks as the only options. However, it generally includes several different implementations of fusion and head networks. Users can simply include the model name as a command line parameter to choose target multi-modal implementation. Besides, \modelname abstracts the training and inference process and integrates them with profiling tools. Users only need to choose proper options to generate desired metrics.

\modelname targets both servers and edge devices. For servers, the training process and inference process can be done within a single python file. For edge devices, only inference is supported due to limited energy and resources. Models must first be trained on servers. Besides, edge devices such as NVIDIA Jetson series~\cite{nvidiaJason} generally adopts a unified memory architecture where GPUs and CPUs share the same physical memory. In this case, adjusting batch size will not affect the memory usage. For large datasets, \modelname will manually split the datasets and inference on partial datasets to ensure the performability.

%--------------
% -----------

\subsection{Profiling Pipeline}
In addition to the  representative workloads, \modelname also provides a series of profiling tools based on the most commonly used hardware available today (CPU and NVIDIA GPU) to help locate the system drawbacks and make corresponding improvements.
The overall profiling architecture is shown in Figure \ref{fig:profile}.
\modelname provides different command line flags to support different measurements options.
To ensure the authority of the measured results, \modelname measures the performance of the network based on standard tools such as, \textit{Python Memory Profiler}~\cite{memory_profiler}, \textit{Pytorch Profiler}~\cite{pytorch_profiler}, \textit{NVIDIA Nsight Compute}~\cite{nsightcompute} and \textit{NVIDIA Nsight System}~\cite{nsightsys}. To ease the usage, \modelname also automates the profiling process using python and shell scripts. 

The evaluation metrics can be categorized into three main classes based on the profiling tools and granularity. The first category includes the inference logs directly generated from the applications. Taking advantage of python modules, \modelname is able to provide basic algorithm level information such as model accuracy, parameter number and FLOPs. The second category includes the entire system information such as GPU information, CPU information and the data transfer between host and device. The third category includes more fine-grained GPU information such as kernel information and GPU execution stall reasons since GPU is in charge of nearly all the computation.

\section{Evaluation}\label{sec:evaluation}
In this section, we present a detailed evaluation of the proposed MMBench to conduct an in-depth analysis of multi-modal DNNs. 
We first introduce the experimental platforms and prove the effectiveness of our selected applications. We then investigate the characteristics of multi-modal DNNs from three main aspects: multi-stage execution, workload heterogeneity and execution synchronization.

\subsection{Experimental Setups}
While MMBench supports various platforms with CPU and NVIDIA GPU, we conduct the following experiments on a GPU server and two edge devices to demonstrate the utility of our benchmark. The GPU server is equipped with two 2.4 GHz Intel 20-core, Xeon 6148 CPUs and four Nvidia RTX 2080Ti GPUs connected via PCI-e ×16 interface with 11 GB of GDDR6 memory. We use a Jetson Nano with 128-core Maxwell, 4GB LPDDR4 and a Jetson Orin with 2048-core Ampere, 32GB LPDDR5 as our edge devices. 
\begin{figure}[t]
	\centering
	\includegraphics[width=\linewidth]{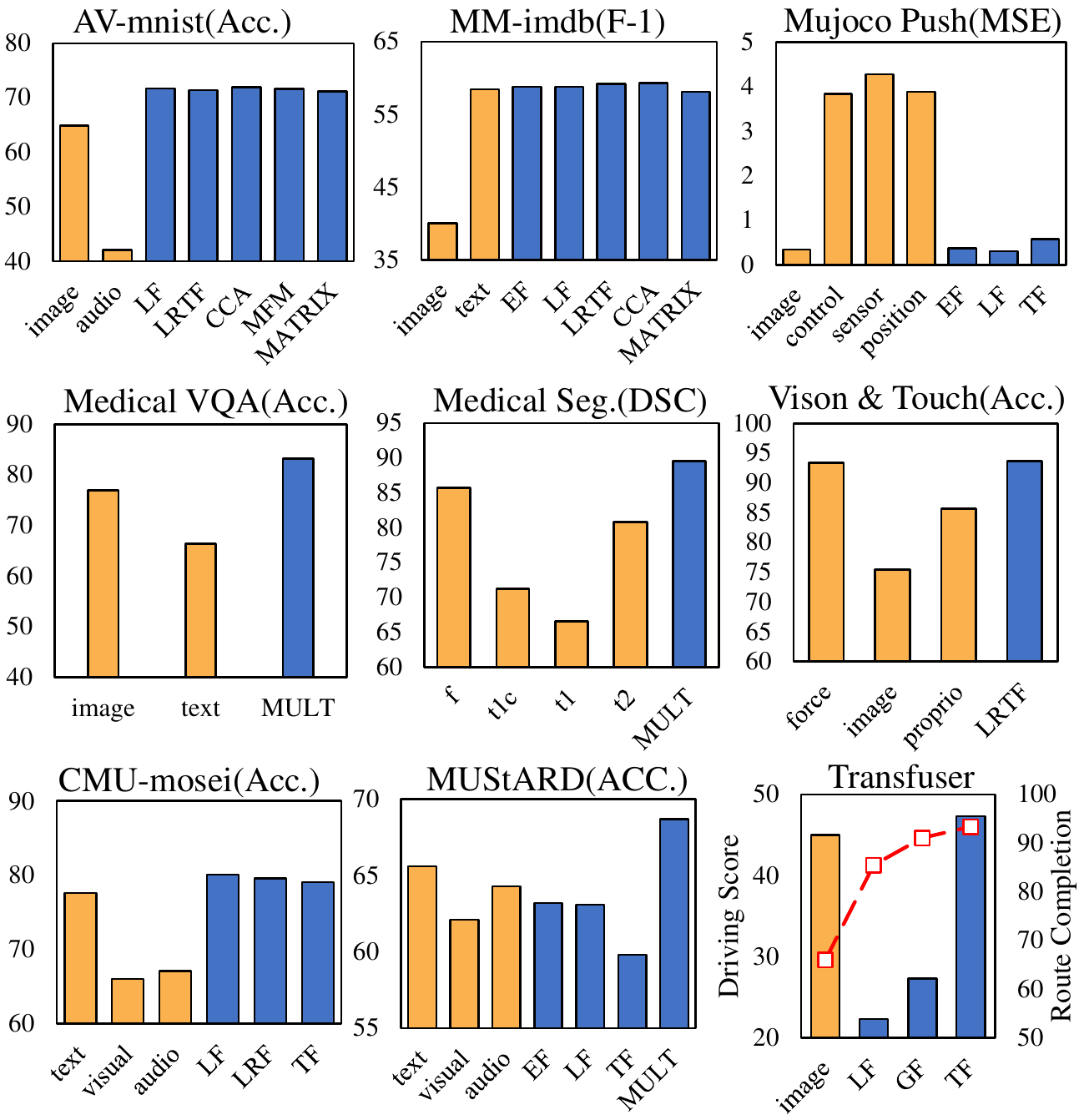}
	\caption{Performance of the applications in \modelnamenospace. Lowercase in yellow such as \textit{image} and \textit{audio} indicates uni-modal implementations, upper case in blue such as \textit{EF} and \textit{LF} indicates multi-modal implementations.} %----------------
	\label{fig:performance}
\end{figure}

\subsection{Network-level Characterization}
\label{sec:accuracy}
In this section, we characterize the algorithm characteristics of multi-modal DNNs. We analyze their overall performance, and the effect of different fusion schemes and modalities in different applications.

\subsubsection{Performance analysis}
We first validate that multi-modal DNNs are able to outperform uni-mdoal DNNs. 
The performance results are presented in different metrics such as accuracy, F-score and MSE. 
To ensure the practical value of \modelnamenospace, we first need to guarantee that all the applications are representative and with high performance. 
Figure~\ref{fig:performance} presents the performance of all the applications included in \modelnamenospace. 
For applications with multiple fusion implementations, we only present several results. For \textit{Transfuser}, multiple metrics such as driving score and route completion are used to evaluate its performance. And the \textit{lidar} modality is seldom executed without \textit{image} modality.
With specific adjustment and optimization, some of the performance can be further improved.

\textit{\textbf{Observations:}
Multi-modal DNNs are proved to outperform uni-modal DNNs in different scenarios. However, multi-modal DNNs generally have various implementations yielding different results. They should be well studied to fully grasp their performance advantage. 
}

\subsubsection{Fusion analysis} 
Most of the current multi-modal researches employ pretrained
DNN backbone models as modality encoders and
focus on finding more effective fusion or representation
methods of different modalities. We examine the influence of different fusion methods on the application performance with same encoders.

Figure~\ref{fig:performance} also presents different fusion implementations for the datasets. Take \textit{Mujoco Push} as example, the MSE of its implementation in late fusion utilizing \textit{LSTM} is less than 0.3 while the MSE of its implementation in \textit{tensor fusion} reaches 0.58. Similarly, in \textit{MM-imdb}, the maximum performance difference between different fusion schemes can be as large as 1.1 in Micro F1.  Some ineffective fusion schemes even lead to lower performance compared with only leveraging single modality. 
The choice of fusion schemes can lead to significant performance variance.

\textit{\textbf{Observations:}
While not significantly influencing the amount of computation for most of the scenarios, different fusion schemes can lead to several percents of absolute performance variance. It's of great importance to design or search for the most effective fusion method.
}

\begin{figure}[t]
    \centering
    \includegraphics[width=\linewidth]{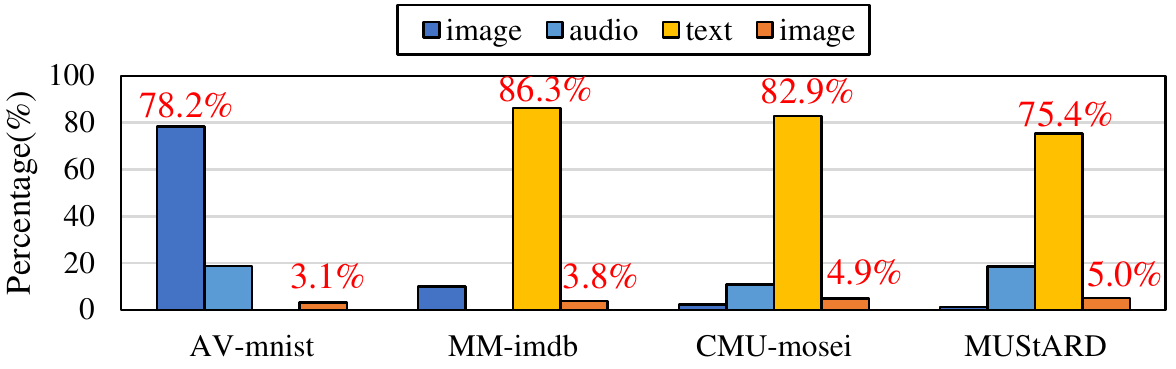}
    \caption{Distribution of mutually exclusive data sample sets correctly processed by different modalities.}
    \label{fig:corect_ratio}
        %\vspace{-5mm}
\end{figure}

\subsubsection{Modality analysis}
In many real-world scenarios, the importance of different modalities differs
depending on the tasks and it is feasible to skip or discard some modality features. 
Typically, some modalities provide higher accuracy with less computational effort than others in different applications. 
For example, it has been proven that text-based features perform better than visual or auditory modalities in  multi-modal language-emotion analysis tasks~\cite{akhtar2019multi}. 

We present the distribution of mutually exclusive data sample sets correctly processed by different modalities in Figure~\ref{fig:corect_ratio}. For the four selected datasets, more than 75\% of the correct samples can be processed using only a major modality while the major modality differs on different tasks. Only less than 5\% of all the correct samples are required to be processed by the multi-modal fusion methods. 
Therefore, one can only rely on some of the encoders given certain tasks to reduce model complexity. 
Under such circumstances, intuitively we can simply throttle sensors for less crucial modalities to save energy. 
However, applying this conventional wisdom is ineffective since it can lead to avoidable task failures resulting from the loss of situation awareness. 
There exists no  retrieval for the extreme conditions where failures occur.

\textit{\textbf{Observations:} Different modalities possess different level of importance in multi-modal DNNs. Smartly activating one of the encoders can fulfill the requirements in most of the cases. There exists room for adaptive execution strategies to achieve a better performance-complexity tradeoff according to the application-specific characteristics. }

\begin{figure}[t]
    \centering
    \includegraphics[width=\linewidth]{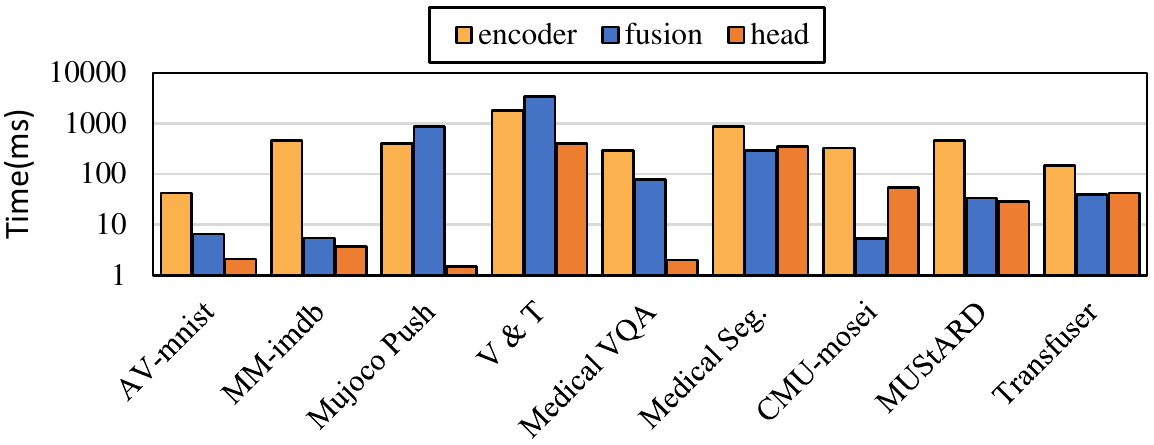}
    \caption{Execution time of a batch of data of the three stages for different \modelname applications.}
    \label{fig:time_stage}
        %\vspace{-5mm}
\end{figure}

\begin{figure*}[t]
    \centering
    \includegraphics[width=\linewidth]{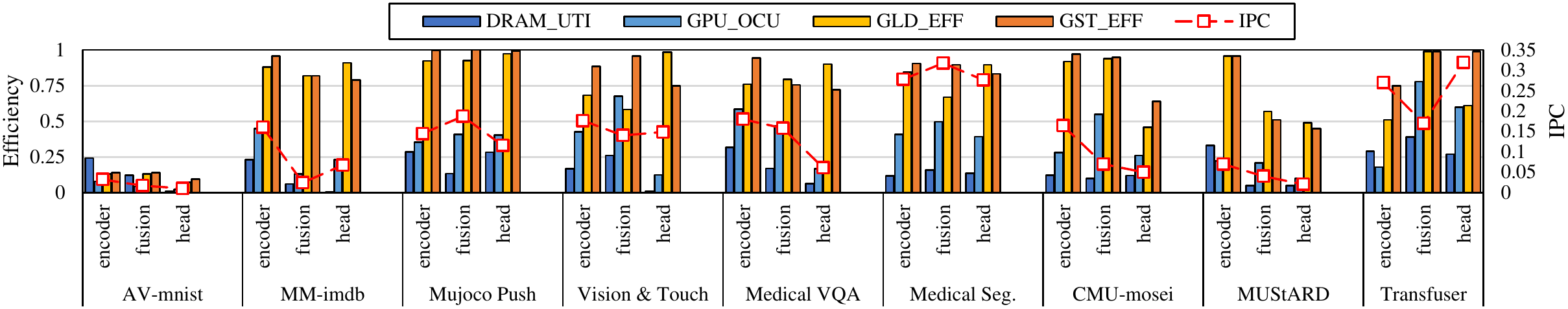}
    \caption{Resource usage of the three stages for different \modelname applications.}
    \label{fig:resource_stage}
        %\vspace{-5mm}
\end{figure*}
\begin{figure*}[t]
    \centering
    \includegraphics[width=\linewidth]{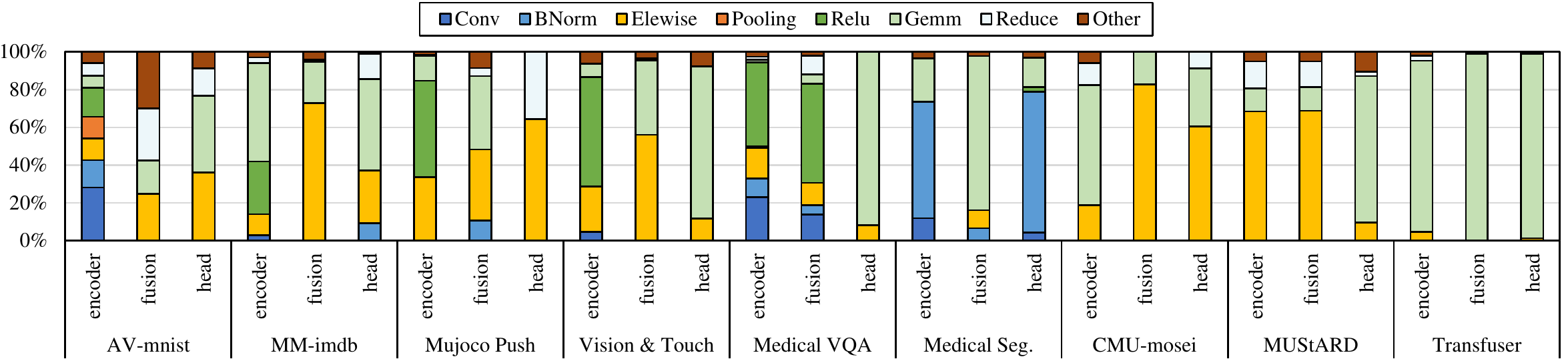}
    \caption{Kernel operation breakdown of the three stages for different \modelname applications.}
    \label{fig:hetero}
        %\vspace{-5mm}
\end{figure*}

\subsection{System-/Architecture-level Analysis}
In this section, we analyze the system and architecture characteristics of multi-modal DNNs. We analyze them according to their three-stage execution pattern, intra-network heterogeneity and frequent synchronization.

\subsubsection{Stage Analysis}
\label{sec:stage_implic}
As introduced in Section~\ref{sec:background}, most multi-modal DNNs can be divided into encoder, fusion and head stages. In this section, we analyze the three-stage execution pattern of multi-modal DNNs. We perform the stage analysis by modifying the forward function. We first record the time consumption of the three stages of the datasets, and investigate the resource usage pattern of different stages. 
Figure~\ref{fig:time_stage} presents the execution time of the applications. 
The execution time distribution depends on specific encoder, fusion and head DNN structures. 
Generally, encoder stage takes much longer time compared with fusion and head stages.
This is because the fusion network takes the learned feature as
input, thus having much smaller data size to deal with.
However, for complex fusion schemes such as \textit{transformer fusion} in the case of \textit{Mujoco Push} and \textit{Vision \& Touch}, it can take even longer time compared with the encoder stage.

We then analyze the resource usage pattern of the \modelname applications in different stages. We trace 5 micro-architectural metrics with \textit{nsight compute}~\cite{nvprof_nvidia}, including \textit{DRAM utilization (1)}, \textit{achieved occupancy (2)}, \textit{ipc (3)}, \textit{gld efficiency} (4) and \textit{gst efficiency (5)}. 
The detailed results are presented in Figure~\ref{fig:resource_stage}. Generally, the encoder stages present higher \textit{DRAM utilization}, \textit{IPC} and \textit{GPU occupancy} compared with fusion and head stages since they include more computation. For \textit{gld efficiency} and \textit{gst efficiency}, all the stages presents nearly the same resource usage pattern. For complex fusion schemes such as \textit{transformer fusion} in \textit{Mujoco Push}, although it takes nearly 3$\times$ more execution time compared with the encoder stage, it does not consume much more resources. While it shows slight increase in \textit{IPC} and \textit{GPU occupancy}, the \textit{DRAM utilization} of the encoder stage is still higher.

\textit{\textbf{Observations:} There exists significant time and resource imbalance in different stages, which leads to possible resource under-utilization. If we assign fixed resources to a multi-DNN application according to its encoder stage, more than half of the resources, especially memory, may actually stay idle when the application enters the fusion and head stages. }

\subsubsection{Heterogeneity Analysis}
Each of the multi-modal encoder sub-network and the fusion network of a multi-modal DNN approximates an independent uni-modal network. Therefore, there is a high degree of heterogeneity within the multi-modal DNNs. In this section, we first analyze the GPU kernel type  breakdown of multi-modal DNNs. We then delve further to analyze the kernel-level information of some hotspot kernels. 
We only select one of the implementations for all the applications.

Figure~\ref{fig:hetero} presents the GPU kernel type breakdown for the applications in \modelnamenospace. We classify all the GPU operations into 8 categories including convolutions (\textit{Conv}), batch normalization (\textit{BNorm}), element-wise operation (\textit{Elewise}), pooling (\textit{Pooling}), relu activation (\textit{Relu}), general matrix multiply (\textit{Gemm}), reduce (\textit{Reduce}) and else (\textit{Other}). 
In this regard, each kernel type contains a subset of function calls that execute similar tasks. 
We observe that different stages within a same application are dominated by different type of operations, not to mention the difference between different applications. 
Besides, the encoder networks for different modalities are highly diverse. 
Some applications, such as \textit{AV-mnist}, apply same encoder network (Lenet) for both modalities. However, \textit{MM-imdb} apply \textit{VGG} and \textit{Albert} to encoder the different modalities. While \textit{VGG} is dominated by \textit{Gemm} (72\%), \textit{Albert} is dominated by \textit{relu} (66\%). Different acceleration strategies are required for these two encoders.

\begin{figure}[t]
	\subfloat[Hotspot kernel: \textit{Reduce} in different stages]{	\includegraphics[width=\linewidth]{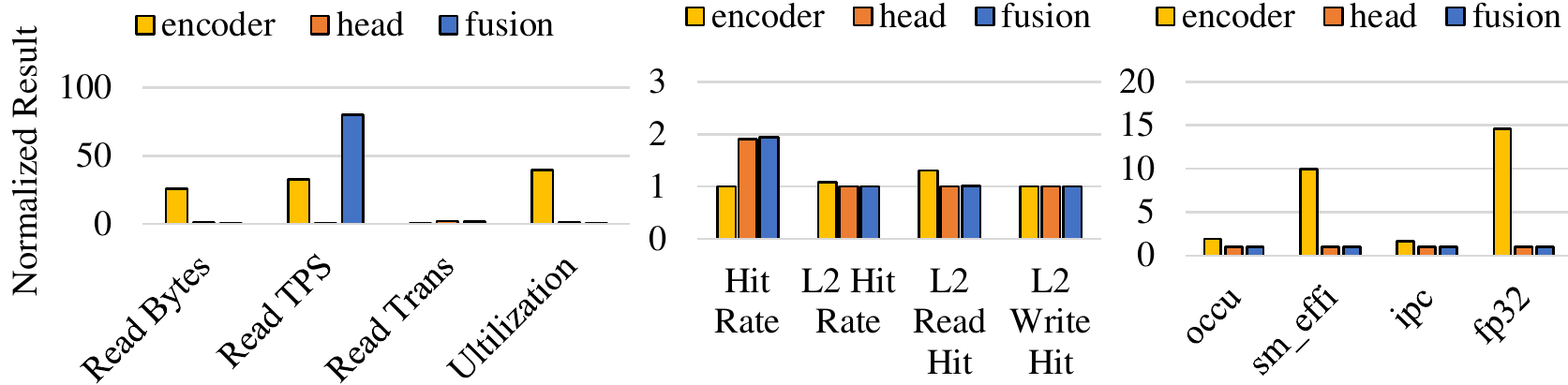}}\\
	\subfloat[Hotspot kernel: \textit{Element-wise} in different fusion methods]{	\includegraphics[width=\linewidth]{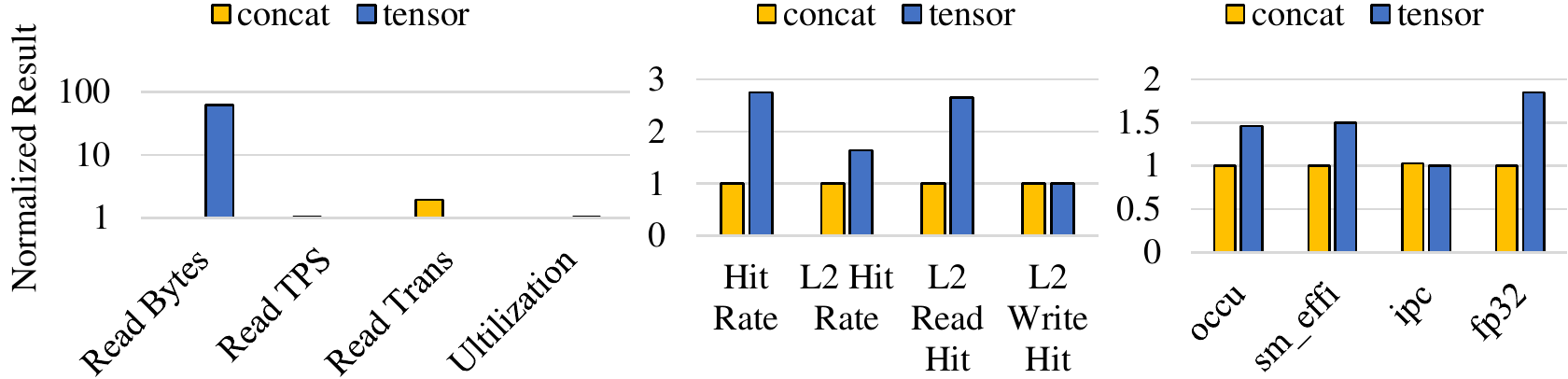}}\\
		\caption{Dedicated kernel comparison different stages and fusion methods on \textit{AV-mnist}. The result is normalized. }\label{fig:uni_slfs_kernel}
\end{figure}

We further choose two hotspot kernels in the case of \textit{AV-mnist} and analyze their fine-grained performance in Figure~\ref{fig:uni_slfs_kernel}. We study the computation, cache and memory patterns of two specific GPU kernels in different stages and different fusion implementations. 
The resource usage of the same kernel in different fusion methods is basically at the same level despite a significant increase in DRAM read bytes. 
However, when it comes to the same kernel in different stages, its average resource usage can vary from 15$\times$ in the total number of \textit{fp32} operations to 80$\times$ in \textit{read TPS}. 
The large difference in memory and compute resources possibly results from the input data size, since fusion and head only handle the learned representations from the encoder stage.

\textit{\textbf{Observations:} There exists different dominant operations in different subnets, and the same operations may perform differently in different stages. In this regard, it is hard to find a universal optimization for the whole multi-modal application. Multi-modal applications must be analyzed first to identify the bottlenecks. It is hard to design specialized hardware accelerators for multi-modal DNN applications.  
}

\subsubsection{Synchronization Analysis}
In this section, we analyze the synchronization problem of multi-modal DNNs. 
From application level, there exists the problem of modality synchronization. The fusion stage must wait until the completion of all modalities.
From network level, multi-modal DNNs suffer from data synchronization. There exists additional intermediate data and data preparation operations which can even overweight GPU computation. 
We first record the execution time of different modalities and then investigate the proportion of CPU+Runtime/ GPU execution.

\begin{figure}[t]
    \centering
    \includegraphics[width=0.95\linewidth]{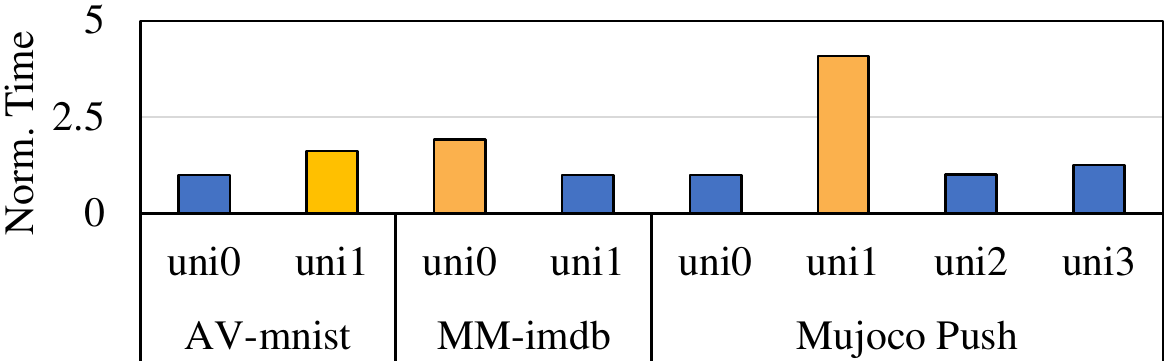}
    \caption{Execution time for different modalities for \modelname applications.}
    \label{fig:uni_time}
        %\vspace{-5mm}
\end{figure}

\textbf{Modality synchronization.} We record the execution time for different modalities.  In Figure \ref{fig:uni_time}, it's obvious that the execution time of different modalities are different. This problem is especially usual for multi-modal tasks involving image modality since image modality generally produces larger amounts of data and require more computation.
For example, the straggler (\textit{uni2: image}) modality in \textit{Mujoco Push} takes up to 4.09$\times$ of inference time compared to other modalities. If executed concurrently,  nearly 75\% of the resources assigned to the application will stay idle for more 77\% of the entire encoder execution.

\begin{figure}[t]
    \centering
    \includegraphics[width=\linewidth]{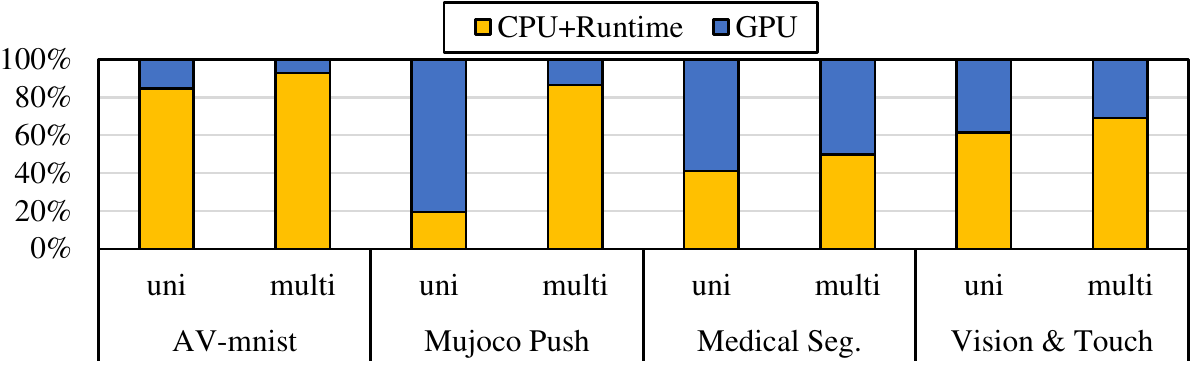}
    \caption{Comparison between synchronization and computation for \modelname applications.}
    \label{fig:data_sync}
        %\vspace{-5mm}
\end{figure}

\textbf{Data synchronization.} Most of the network computation are executed on GPUs, and CPUs are mainly in charge of data processing operations, such as \textit{to} and \textit{copy}.
Thus we consider that a higher \textit{CPU+Runtime} ratio indicates more data synchronization operations as GPU are more frequently kept stalled for lack of data.  
We choose several applications in different research domains to investigate their inference time breakdown. The detailed results are shown in Figure~\ref{fig:data_sync}. We can observe that for all applications, the multi-modal implementation possess larger proportion of \textit{CPU+Runtime} operations compared with the uni-modal implementations.
Complex fusion such as \textit{Mujoco push} can lead to a significant increase in \textit{CPU+Runtime} of 66\%.


\textit{\textbf{Observations:}  
Multi-modal DNNs suffer from two-level of synchronization. From application level, its encoder subnets requires modality synchronization before the fusion stage. The fusion network must always wait for the straggler.
From operator level, lengthy intermediate data operations lead to frequent data synchronization. 
These altogether leads to the resource under-utilization problem, as GPUs may stay idle for most of the application time. }

\section{Two Case Studies Using \modelnamenospace} \label{sec:case_study}
\subsection{Effect of Batch Size}
Beyond simply executing the multi-modal DNN applications, MMBench also provides multiple tuning knobs to study the effect of different parameters and help adjust the system. Here we provide a case study, demonstrating how MMbench help explore the effect of batch size on multi-modal DNNs compared with uni-modal DNNs. 

\begin{figure}[t]
   \includegraphics[width=\linewidth]{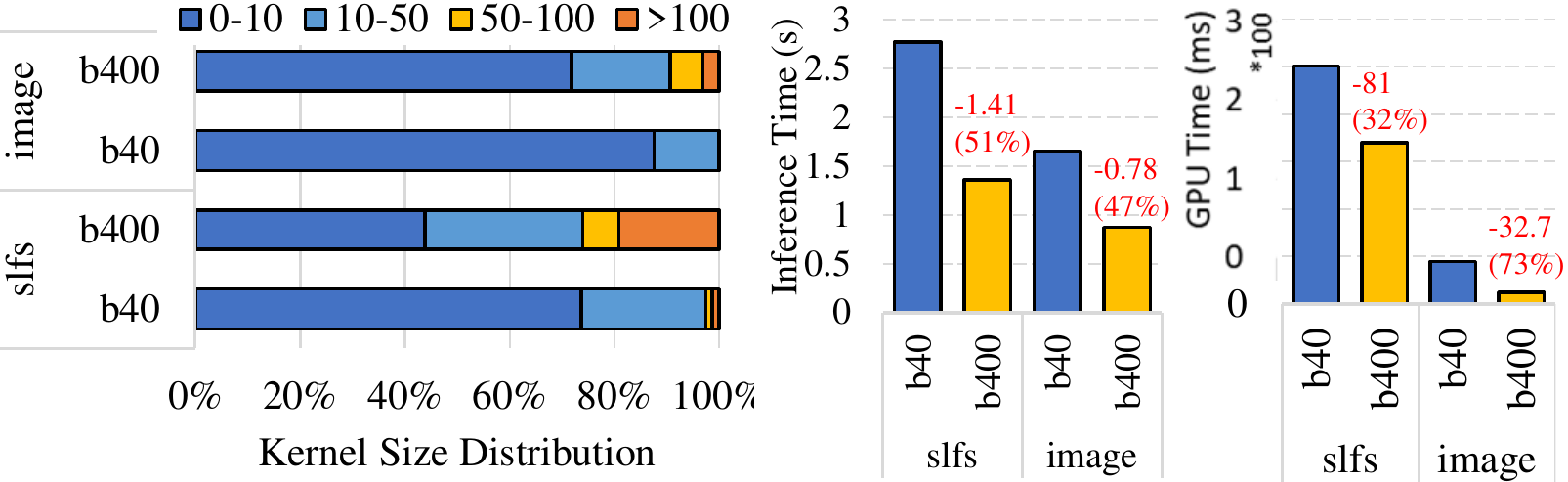}
	% \subfloat[Comparison of different stages]{			\includegraphics[width=\linewidth]{batch_stage_all.png}}	\\
	\vspace{2mm}
		\caption{Larger batch size can accelerate the execution of multi-modal DNNs on \textit{AV-mnist}. \textit{Slfs} is a multi-modal implementation, and \textit{image} is the uni-modal counterpart. \textit{b} refers to batch size.}\label{fig:batch_scheduling}
\end{figure}
\begin{figure}[t]
    \subfloat[Uni-modal DNN]{			\includegraphics[width=0.48\linewidth]{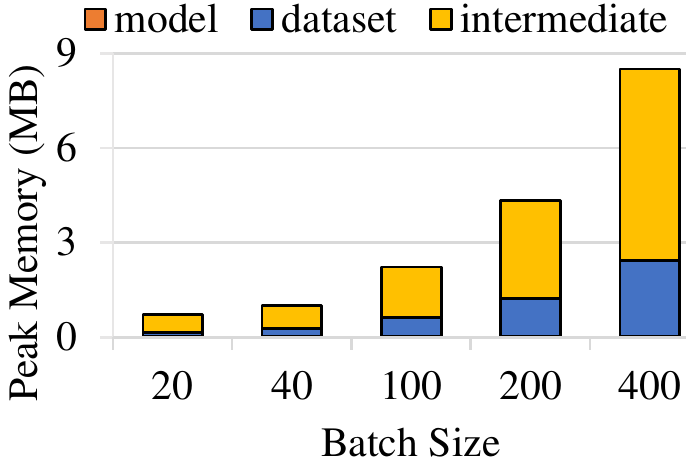}}
	\subfloat[Multi-modal DNN]{			\includegraphics[width=0.51\linewidth]{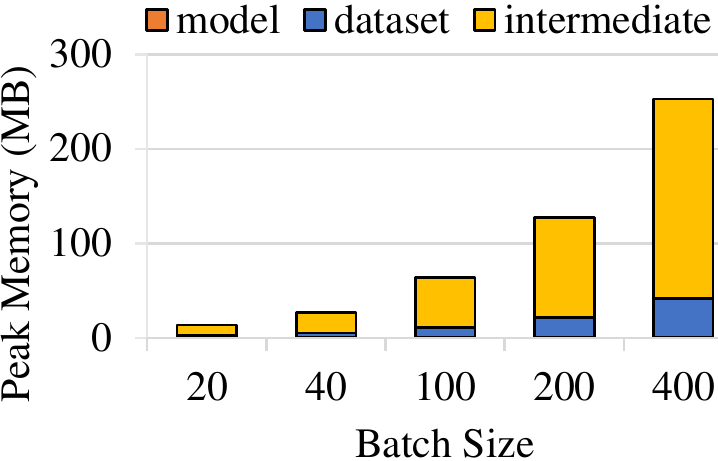}}	\\
		\caption{Peak memory for processing models, datasets, and intermediate results on \textit{AV-mnist}.}\label{fig:peak_memory}
\end{figure}
% -------------------
Generally, when a batch of tasks arrive, the operating system schedules the appropriate kernels to handle those tasks. If we ignore the computational differences among various kernels of different sizes, a batch size of 400 tasks will be executed in 10$\times$ less time than a batch of 40 tasks. However, this is impossible during real execution due to resource contention and constraints.   
The current OS often leverages larger kernels which yield better tradeoff between GPU time and non-GPU time (e.g., data transfer time, synchronization time etc.) to process a large batch of tasks. In this regard, it is more beneficial to process multi-modal DNN tasks in large batch size. Figure \ref{fig:batch_scheduling} shows our analysis. We consider 10000 inference tasks which are scheduled with batch size of 40 and 400 respectively. 

We first analyze the distribution of kernels of different sizes. Based on the GPU execution time of each kernel, we divide the kernels into four different kernel sizes. Figure \ref{fig:batch_scheduling} illustrates the comparison results of uni-modal and multi-modal DNNs. \textit{0-10} indicates a small kernel, where the kernel executes in less than 10 microseconds. \textit{>100} indicates a large kernel, where the kernel executes in more than 100 microseconds. The leftmost result shows that existing operating system (OS) uses more large kernels whose execution time exceed 50 microseconds to process a larger batch size with 400 tasks. Meanwhile, the OS calls more large kernels to process the multi-modal DNN tasks. The results on the right show that a 10x increase in batch task size does not reduce the processing latency by 10x.

Besides, as shown in Figure \ref{fig:peak_memory}, larger batch size leads to higher peak memory usage for model, dataset and intermediate features. The model sizes remain generally the same, while the dataset and intermediate features present a linear order relative to batch size. 
Multi-modal DNNs also tend to produce higher proportion of intermediate data. 
When changing batch size, multi-modal DNNs are easier to achieve GPU memory capacity since they involve more intermediate features with multiple modalities and additional fusion networks.

\textit{\textbf{Observations:} 
Our analysis shows that different components of multi-modal DNNs benefits differently from batch size. 
The GPU time of multi-modal DNNs decrease in a smaller scope, which possibly results from the kernel composition of the networks.
Besides, batch size increase leads to higher growth rate in peak memory usage for multi-modal DNNs.}

\begin{figure}[t]
\centering
    
  	\includegraphics[width=\linewidth]{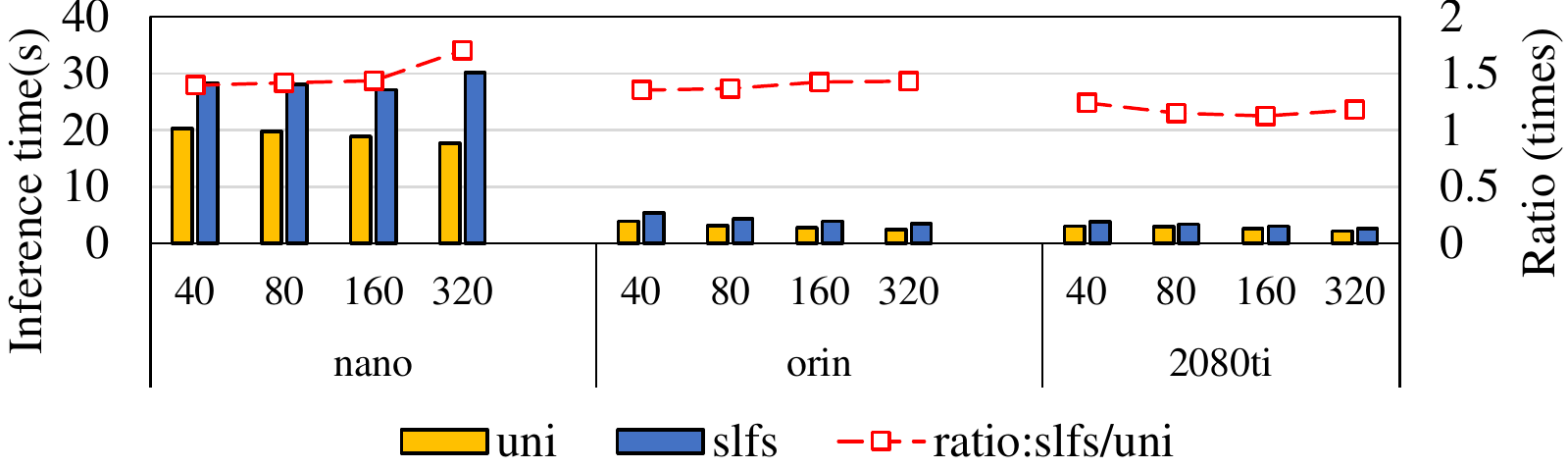}
  	\caption{Inference time of \textit{AV-mnist} on GPU server and edge devices with the change of batch size. \textit{slfs} refers to an implementation of multi-modal with 31x parameters. \label{edge-inference}}
\end{figure}

\subsection{Migration to Edge Computing} \label{sec:edge}

In recent years, there has been an increasing trend to deploy DNN models at the edge due to the connectivity, latency and privacy concerns of transferring data to the cloud. 
Therefore, we also characterize the features of multi-modal DNNs at the edge. We run \textit{AV-mnist} on one of the most representative AIoT boards, i.e., Jetson Nano.

\begin{figure}[t]
    \subfloat[Breakdown of stall cycles on Jetson Nano]{			\includegraphics[width=\linewidth]{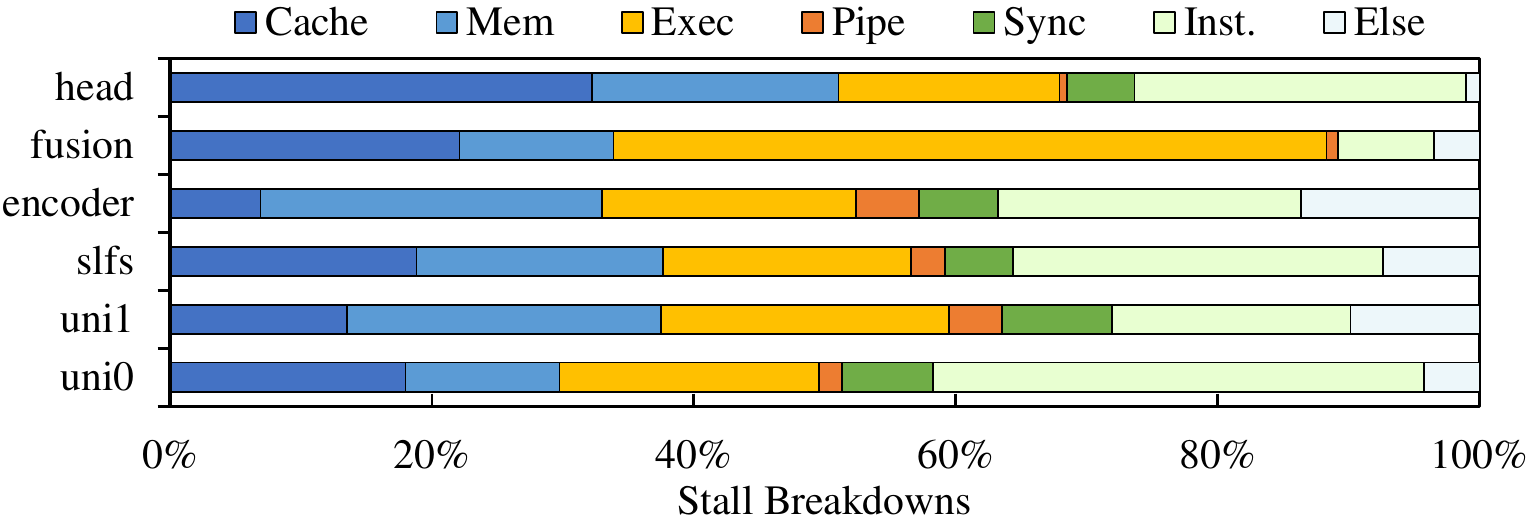}}\\
    \subfloat[Breakdown of stall cycles on 2080ti]{			\includegraphics[width=\linewidth]{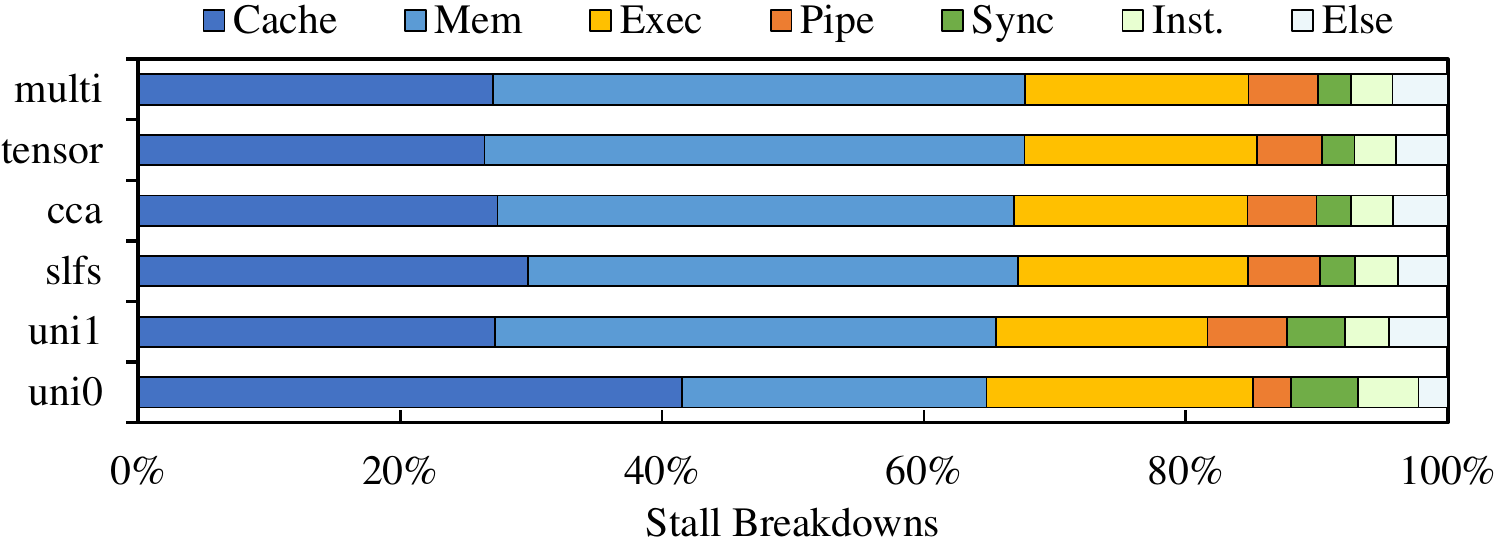}}\\
	\subfloat[Computation and memory usage on Jetson Nano]{			\includegraphics[width=\linewidth]{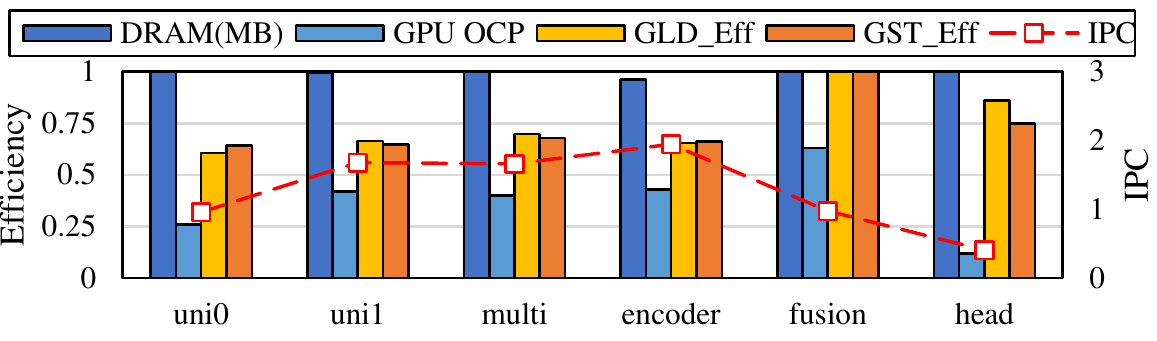}}	\\
		\caption{Execution stall breakdown and resource usage on edge devices. \textit{uni0} refers to audio, \textit{uni1} refers to image.}\label{fig:nano_result}
		\vspace{-2mm}
\end{figure}

Figure~\ref{edge-inference} presents the inference time of \textit{AV-mnist} on both GPU servers and edge devices. On Jetson nano where resources are limited, 6.48$\times$ more time is needed compared with GPU servers.
With the increase of batch size, while the latency of GPU server is constantly decreasing, the latency of Jetson nano is even higher when batch size reaches 320. It is because certain resources are used up. On Jetson orin with abundant resources, the multi-modal DNNs perform similarly as on GPU servers. The ratio of multi-modal execution time compared with uni-modal is higher on Jetson nano and orin, since GPU servers possess more idle resources.

In Figure \ref{fig:nano_result}-(a) and (b), we illustrate the execution stall breakdown and resource usage patterns of multi-modal DNN both on edge devices and on GPU servers.
We divide the stall reasons into 7 main categories: 
\textit{cache dependency (Cache)}, \textit{memory dependency (Mem)}, \textit{execution dependency (Exec)}, \textit{busy pipeline (Pipe)}, \textit{synchronization blocked (Sysn)}, \textit{instruction not fetched (Inst.), \textit{other stalls (Else)}}.
The stall caused by \textit{execution dependency} and \textit{instruction not fetch} increases dramatically on edge devices, while \textit{memory dependency} and \textit{cache miss} are the main causes of stall on GPU servers. 
It possibly results from the lack of computing power so that requisite operations cannot be finished in time.
As shown in Figure \ref{fig:nano_result}-(c), on edge devices with limited resources, \textit{DRAM utilization} is almost always kept at the highest level. 
Unlike GPU servers in Figure \ref{fig:resource_stage}, fusion stage now possesses higher \textit{GPU occupancy} on edge devices.

\textit{\textbf{Observations:} Migration to edge devices leads to higher latency and new bottlenecks. Due to limited power and resources, the inference time grows dramatically when we switch from uni-modal DNN to multi-modal DNNs even on small datasets. It would be a huge challenge to enable multi-modal DNNs on edge devices. Some of the modalities may be skipped to guarantee the QoS on edge devices as long as the result meets the requirement.}

\section{Conclusion}\label{sec:conclusion}
We present \modelnamenospace, an open-source benchmark suite for end-to-end cloud and IoT multi-modal neural networks. The suite includes multiple representative multi-modal computing applications, such as multimedia analysis, affective computing, medical analysis, etc. We use \modelname to study the system and architectural implications of multi-modal neural networks across different computing stacks and conclude three unique characteristics. We also provide two case studies to demonstrate how \modelname guides the system and architecture designs. We expect that our work could pave the way for better system and architecture research for multi-modal computing. 

\section*{Acknowledgments}

This work is supported  by the National Natural Science Foundation of China (No.62122053). Corresponding authors are Xiaofeng Hou and Chao Li. We thank all the anonymous reviewers for their
valuable comments and suggestions.

\bibliographystyle{IEEEtranS}
\bibliography{IEEEexample}

\end{document}